 \documentclass[pra,superscriptaddress,showpacs,twocolumn]{revtex4}

\usepackage{epsfig,graphics}

\def \E{{\rm e}}
\def \tr{{\rm Tr}\,}
\def \>{\rangle}
\def \<{\langle}

\begin{document}
\title{Optimal mirror phase-covariant cloning}

\author{Karol Bartkiewicz} 
\affiliation{Faculty of Physics, Adam Mickiewicz University,
61-614 Pozna\'n, Poland}

\author{Adam Miranowicz}
\affiliation{Faculty of Physics, Adam Mickiewicz University,
61-614 Pozna\'n, Poland} \affiliation{Graduate School of
Engineering Science, Osaka University, Osaka 560-8531, Japan}

\author{\c{S}ahin Kaya \"Ozdemir}
\affiliation{Graduate School of Engineering Science, Osaka
University, Osaka 560-8531, Japan} \affiliation{Department of
Electrical and Systems Engineering, Washington University in St.
Louis, St. Louis, Missouri 63130, USA}

\begin{abstract}
We propose a quantum cloning machine, which clones a qubit into
two clones assuming known modulus of expectation value of Pauli
$\sigma_z$-matrix. The process is referred to as the mirror
phase-covariant cloning, for which the input state is {\em a
priori} less known than that for the standard phase-covariant
cloning. Analytical expressions describing the cloning
transformation and fidelity of the clones are found. Extremal
equations for the optimal cloning are derived and analytically
solved by generalizing a method of Fiur\'a\v{s}ek [Phys. Rev. A
{\bf 64}, 062310 (2001)]. Quantum circuits implementing the
optimal cloning transformation and their physical realization in a
quantum-dot system are described.
\end{abstract}
\date{\today}
\pagestyle{plain} \pagenumbering{arabic}

 \maketitle

\section*{Introduction}

One of the fundamental no-go theorems in quantum mechanics is the
no-cloning theorem \cite{Zurek}, which states that unknown quantum
states cannot be perfectly copied. In other words, no quantum
mechanical evolution exists which would transform a quantum state
according to $|\phi\>\rightarrow |\phi\>|\phi\>$ for an unknown
state $|\phi\>$. This theorem is a consequence of the linearity of
quantum mechanics, and it has tremendous technological
implications; e.g., it is at the basis of the security of quantum
communication protocols including quantum key distribution.
Although exact cloning is impossible, pretty good approximate
cloning is possible as shown for the first time by Bu\v{z}ek and
Hillery \cite{Buzek96}. They designed a cloning machine, referred
to as the $1\rightarrow 2$ {\em universal cloner} (UC), which
produces two approximate copies from an unknown pure qubit state.
The UC is a state-independent symmetric cloner in the sense that
all qubit states are cloned with the same fidelity $F=5/6$ and
fidelities of the clones to the initial pure state is the same
$F_1=F_2$. The concept of cloning has attracted considerable
interest, and it has been later shown that for the $1\rightarrow
M$ UC, the relation between the optimum fidelity $F$ of each copy
and the number $M$ of copies is given by $F=(2M+1)/(3M)$
\cite{Gisin}. Setting $M\rightarrow\infty$ corresponds to a
classical cloning machine with $F=2/3$, which is the best fidelity
that one can achieve with only classical operations. Moreover, the
concept has been extended to include cloning of qudits, cloning of
continuous-variable systems or state-dependent cloning
(nonuniversal cloning), which can produce clones of a specific
set of qubits with much higher fidelity than the rest
\cite{Buzek98,Bruss98,Bruss00,F1,F2,Fan,Buscemi,Sacchi,Kay} (for
reviews see \cite{cloning}). This paper is devoted to the latter
topic.

Suppose we want to clone a qubit, which is in a pure state
\begin{equation}
|\psi\>=\cos{\frac{\vartheta}{2}}|0\>+\E^{i\phi}\sin{\frac{\vartheta}{2}}|1\>
\label{N01}
\end{equation}
parametrized by polar $\vartheta$ and azimuthal $\phi$ angles on
the Bloch sphere. By considering only the $1\rightarrow 2$
cloning, the joint density matrix of both clones can be given by
\cite{F1,F2}:
\begin{eqnarray}
\rho_{\rm out}&=&\tr_{{\rm in}}\left(\chi\rho_{{\rm in}}^{{\rm T}}
\otimes \openone_{\rm out}\right),
 \label{N02}
\end{eqnarray}
where $\chi$ is a trace-preserving completely positive (TPCP) map
describing the cloning operation in the tensor product of the
input (${\cal H}_{\rm in}$) and output (${\cal H}_{\rm out}$)
Hilbert spaces. Moreover, $\tr_{{\rm in}}$ stands for partial
trace over ${\cal H}_{\rm in}$, $\rho_{{\rm in}}=|\psi\>\<\psi|$,
$T$ denotes transposition, and $\openone_{\rm out}$ is the
identity operator in ${\cal H}_{\rm out}$. The quality of the
cloning can be described by the single-clone fidelity
\begin{eqnarray}
F_{j}(\vartheta,\phi)&=& \<\psi|\rho_{j}|\psi\>,
 \label{N03}
\end{eqnarray}
where $\rho_{j}=\tr_{j\oplus1}\left(\rho_{\rm out}\right)$ is the
reduced density matrix of the $j$th clone ($j=1,2$). As shown in
\cite{F1}, the map $\chi$ can be found by using an optimization
procedure which maximizes the fidelity of the clones. In order to
find a map $\chi$, which makes clones of pure state qubits of the
best possible quality using a partial knowledge about the states,
one needs to maximize the average single-copy fidelity
\begin{equation}
F=\frac12\int_0^{2\pi} {\rm d}\phi \int_0^\pi {\rm d}\vartheta\, g
(\vartheta,\phi) [F_1(\vartheta,\phi)+F_2(\vartheta,\phi)],
\label{N04}
\end{equation}
which is an average over all possible input qubits defined by the
distribution function $g (\vartheta,\phi)$.

The study of state-dependent cloning machines is important as it
is often the case that we have some {\it a priori} information on
the quantum state but we do not known it exactly. Using the
available {\it a priori} information, we can then design a cloning
machine which performs better cloning than the UC for some
specific set of qubits. For example, if it is known that the qubit
is chosen from the equator of the Bloch sphere, then we know that
$\phi$ can be arbitrary while $\vartheta=\pi/2$. For such a case,
the so-called {\em phase-covariant cloners} (PCCs) have been
designed \cite{Bruss00,Fan}, and they have been shown to be
optimal providing a higher fidelity than the UC. Fiur\'a\v{s}ek
\cite{F1,F2} studied the PCCs with known $\vartheta=\theta$ from
the full range $[0,\pi]$ and provided two optimal symmetric
cloners; one for the states in the lower and the other for those
in the upper hemisphere of the Bloch sphere.

In this paper, we assume less {\it a priori} information and
construct an optimal $1\rightarrow 2$ symmetric cloner using the
approach developed by Fiur\'a\v{s}ek \cite{F1,F2}. Contrary to the
works of Fiur\'a\v{s}ek, where the cloners are designed for known
fixed value of $\theta$, that is fixed $\sigma_z$ component, we
provide an optimal cloner for the qubits with known $\sin\theta$
(or, equivalently, $|\<\sigma_z\>|$). Thus, our cloner prepares
two symmetric clones for any qubit with $\vartheta=\theta$ or
$\pi-\theta$.

\section{Optimal cloning and partial knowledge about initial state}

\begin{figure}[t!]
\hspace*{-3mm} \epsfxsize=6cm
\epsfbox{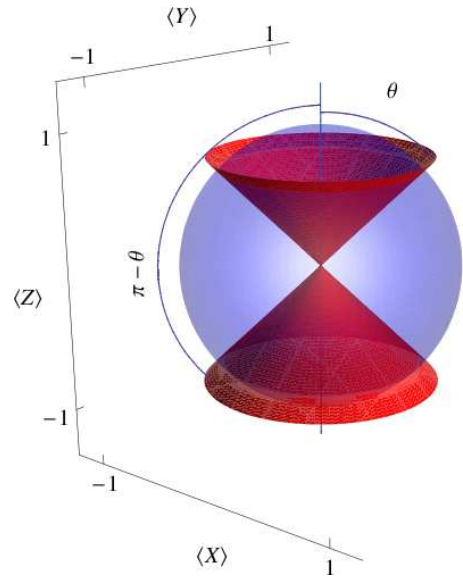}\hspace*{0mm}\epsfxsize=4cm \vspace{3mm}
\caption[]{(Color online) The intersection of two cones and a
Bloch sphere provides the set of qubits that we want to clone in
an optimal way. Here, $X,Y,Z$ denote Pauli operators.}\label{fig1}
\end{figure}

Suppose that we are given a qubit prepared in a pure state
$|\psi\>$, given by Eq. (\ref{N01}), together with the expectation
value of the observable $\sigma_z$, that is $\phi$ is arbitrary
but $\vartheta$ is equal either to $\theta$ or to $\pi-\theta$.
Our task is to find a $1\rightarrow 2$ cloning machine which
prepares optimal approximate clones of the input qubit. The input
qubits of interest are those lying along the intersection of two
cones, which have the cone angles $\theta$ and $\pi-\theta$,
sharing the same apex with Bloch sphere as shown in Fig.
\ref{fig1}. Thus, the qubits are from both the upper and lower
hemispheres of the Bloch sphere. We can consider the qubits from
the lower hemisphere as the mirror images of those from the upper
hemisphere or vice versa. Therefore, we suggest to call this
cloning machine as the {\em mirror phase-covariant cloner} (MPCC)
and require that it satisfies the following conditions: (i) qubits in the
upper and lower hemisphere are cloned with the same maximal
fidelity, $F(\theta)=F(\pi-\theta)$, and (ii) the sum of the
fidelities of the two clones is the maximum attainable fidelity.

To derive the transformation we maximize the following functional:
\begin{equation}
F=\tr\left(\chi R\right),
 \label{N05}
\end{equation}
where $\chi$ is a positive map and $R=\frac12\left(r_{\theta}
+r_{\pi-\theta}\right)$ given in terms of
\begin{eqnarray}
r_{\theta }=\frac{1}{8}{\left(
\begin{array}{cccccccc}
8c_{2}^{4} & 0 & 0 & 0 & 0 & s_{1}^{2} & s_{1}^{2} & 0 \\
0 & 4c_{2}^{2} & 0 & 0 & 0 & 0 & 0 & s_{1}^{2} \\
0 & 0 & 4c_{2}^{2} & 0 & 0 & 0 & 0 & s_{1}^{2} \\
0 & 0 & 0 & 2s_{1}^{2} & 0 & 0 & 0 & 0 \\
0 & 0 & 0 & 0 & 2s_{1}^{2} & 0 & 0 & 0 \\
s_{1}^{2} & 0 & 0 & 0 & 0 & 4s_{2}^{2} & 0 & 0 \\
s_{1}^{2} & 0 & 0 & 0 & 0 & 0 & 4s_{2}^{2} & 0 \\
0 & s_{1}^{2} & s_{1}^{2} & 0 & 0 & 0 & 0 & 8s_{2}^{4}
\end{array}
\right) },
 \label{N06}
\end{eqnarray}
where  $s_i=\sin(\theta/i)$ and $c_i=\cos(\theta/i)$ for $i=1,2$.
The derivation of Eq.~(\ref{N05}) is based on the method described
in Ref. \cite{F2}. For the MPCC, $g$ distribution occurring in
(\ref{N04}) is given by
\begin{equation}
g_{\theta}(\vartheta,\phi) =\frac1{4\pi}[\delta(\vartheta-\theta)
+\delta(\vartheta+\theta-\pi)], \label{N07}
\end{equation}
in terms of Dirac's $\delta$ function. Moreover, subscript $\theta$
was added to indicate {\em a priori} knowledge about the input
state. Note that $g_{\theta}(\vartheta,\phi)$ is equal to
$\frac1{4\pi}\sin\vartheta$ for the UC and to
$\frac1{2\pi}\delta(\vartheta-\theta)$ for the PCC. By defining a
$\phi$-averaged fidelity for each of the clones ($i=1,2$)
\begin{equation}
F_i(\theta)=\frac{1}{2\pi} \int_0^{2\pi} F_i(\theta,\phi) {\rm
d}\phi,
 \label{N08}
\end{equation}
one can rewrite Eq. (\ref{N04}) assuming Eq. (\ref{N07}) as
\begin{equation}
F(\theta) = \frac14 [F_1 (\theta)+F_2
(\theta)+F_1(\pi-\theta)+F_2(\pi-\theta)],
 \label{N09}
\end{equation}
and this quantity is to be maximized. By contrast, $F(\theta) =
\frac12 [F_1 (\theta)+F_2 (\theta)]$ was applied in the
maximization procedure for the PCC. Using the method given in Eq.
\cite{F1} we derive map $\chi$ in the following form:
\begin{eqnarray}
\chi(\theta)={ \left(\begin{array}{cccccccc}
A&0&0&0&0&C&C&0\\
0&B&B&0&0&0&0&C\\
0&B&B&0&0&0&0&C\\
0&0&0&0&0&0&0&0\\
0&0&0&0&0&0&0&0\\
C&0&0&0&0&B&B&0\\
C&0&0&0&0&B&B&0\\
0&C&C&0&0&0&0&A
\end{array}\right),}
\label{N10}
\end{eqnarray}
where $A,B$, and $C$ are $\theta$ dependent. Moreover, $B=(1-A)/2$
as required by trace preservation of the transformation, and
$C=\sqrt{AB}$ as will be shown analytically in the following.

\section{Analytical results}

Using the Kraus decomposition with $\chi$, as described in
\cite{F1}, we find that one of the possible unitary
implementations of the TPCP map can be written as
\begin{eqnarray}\nonumber
|0\>|0\>_{\rm anc}\rightarrow \frac{A|00\>|0\>_{\rm anc}
+\sqrt{2}C|\psi_+\>|1\>_{\rm anc}}{\sqrt{A^2+2C^2}},\\
|1\>|0\>_{\rm anc}\rightarrow \frac{A|11\>|1\>_{\rm
anc}+\sqrt{2}C|\psi_+\>|0\>_{\rm anc}}{\sqrt{A^2+2C^2}},
\label{N12}
\end{eqnarray}
where $|\psi_+\>=\frac{1}{\sqrt{2}}\left(|01\>+|10\>\right)$. This
result is confirmed by our numerical analysis.

Fidelity of the clones can be derived by making use of the unitary
transformation (\ref{N12}) and the explicit form of $|\psi\>$
given by Eq. (\ref{N01}). After performing the calculations we derive
\begin{equation}
F=\frac{1+\Lambda^2}{2}-\frac{1}{2}\sin^2\theta
\left(\Lambda^2-\Lambda\sqrt{2-2\Lambda^2}\right)
 \label{N14}
\end{equation}
where $\Lambda\equiv A/\sqrt{A^2+2C^2}$. For the transformation to
maximize fidelity, i.e., for all $\theta\in[0,\pi]$ we impose
\begin{equation}
\frac{\partial F}{\partial \Lambda}=0
 \label{N13}
\end{equation}
from which we get four expressions for $\Lambda$ ($i,j=0,1$):
\begin{eqnarray}
\Lambda_{i+2j}&=&(-1)^{i}\sqrt{\frac{1}{2}+(-1)^{j}
\frac{\cos^2\theta}{2\sqrt{P}}},
 \label{N15}
\end{eqnarray}
where $P\equiv P(\theta)=2-4\cos^2\theta+3\cos^4\theta$ reaches
the extremum values $P_{\rm min}=P\left[ {\rm
acos}(\sqrt6/3)\right]=P\left[\pi- {\rm acos}
(\sqrt6/3)\right]=2/3$ and $P_{\rm max}=P(\pi/2)=2$. Only one of
the solutions provides fidelity, which is as high as the one
derived numerically. Therefore, $\Lambda$ is given by
\begin{equation}
\Lambda\equiv\Lambda_{0}.
 \label{N16}
\end{equation}
and then
\begin{equation}
\nonumber A=\Lambda^2,\quad B=\frac{1}{2}\bar{\Lambda}^2,\quad
C=\frac{1}{\sqrt{2}}\Lambda\bar{\Lambda}, \label{N17}
\end{equation}
where $\bar{\Lambda}=\sqrt{1-\Lambda^2}$. Now the unitary
transformation can be written as
\begin{eqnarray}\nonumber
|0\>|0\>_{\rm anc}&\rightarrow& \Lambda|00\>|0\>_{\rm anc}+\bar{\Lambda}|\psi_+\>|1\>_{\rm anc},\\
|1\>|0\>_{\rm anc}&\rightarrow& \Lambda|11\>|1\>_{\rm
anc}+\bar{\Lambda}|\psi_+\>|0\>_{\rm anc}
 \label{N18}
\end{eqnarray}
leading to the following reduced density matrices of the clones
\begin{eqnarray}
\rho_{1}=\rho_{2}=\left[\begin{array}{cc}
{\frac{1}{2}}(1+\Lambda^{2}\cos\theta), & \frac{1}{\sqrt{2}}{\rm e}^{-i\phi}\Lambda\bar{\Lambda}\sin\theta\\
\frac{1}{\sqrt{2}}{\rm e}^{i\phi}\Lambda\bar{\Lambda}\sin\theta, &
\frac{1}{2}(1-\Lambda^{2}\cos\theta)\end{array}\right].
\label{N101a}
\end{eqnarray}
Thus, fidelity of the clones, $F=F_{1}=F_{2}$, created by this
transformation can easily be calculated as
\begin{equation}
F=\frac{1}{2}(1+\Lambda^{2}\cos^{2}\theta
+\sqrt{2}\Lambda\bar{\Lambda}\sin^{2}\theta).
 \label{N19}
\end{equation}
In Fig. 2, this fidelity is depicted in comparison to fidelities
for the optimal PCC and UC. As seen in Fig.~2, $F(\theta)$ has
two minima $F\left({\rm acos}(\sqrt3/3)\right)=F\left(\pi-{\rm
acos}(\sqrt3/3)\right)=5/6$ and a local maximum
$F\left(\pi/2\right)=1/2+\sqrt2/4$. The eigenstates of Pauli
$\sigma_z$ matrix are cloned with the highest fidelity
$F\left(0\right)=F\left(\pi\right)=1$.

\begin{figure}[t!]
\hspace*{-3mm} \epsfxsize=7cm
\epsfbox{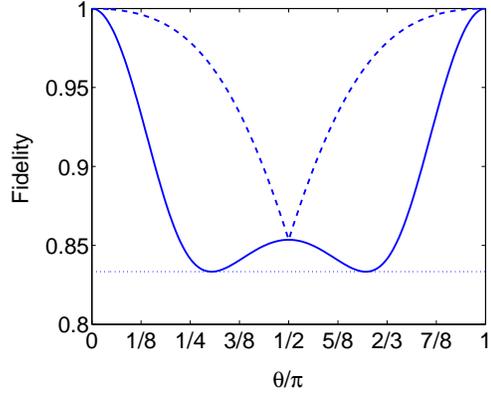}\hspace*{0mm}\epsfxsize=4cm \vspace{3mm}
\caption[]{Comparison of the $\theta$ dependence of fidelities for
the three optimal cloning machines: the MPCC (solid), the PCC
(dashed), and the UC (dotted curve). }\label{fig2}
\end{figure}

For better visualization of the cloned states generated by the
optimal MPCC we depicted their Bloch representation in Fig. 3 in
comparison to the corresponding representations for the other
optimal cloning machines. Specifically, the Bloch vector
$\vec{r}=(\langle\sigma_{x}\rangle,\langle\sigma_{y}
\rangle,\langle\sigma_{z}\rangle)$ of each clone is found to be
\begin{eqnarray}
\vec{r}=[\sqrt{2}\Lambda\bar{\Lambda}\cos\phi\sin\theta,
\sqrt{2}\Lambda\bar{\Lambda}\sin\phi\sin\theta,\Lambda^{2}\cos\theta],
 \label{N103}\end{eqnarray}
which is in contrast to
$\vec{r}=\frac{2}{3}[\cos\phi\sin\theta,\allowbreak
\sin\phi\sin\theta,\allowbreak \cos\theta]$ for the UC and to the
Bloch vector
\begin{equation}
\vec{r}=\left[\frac{1}{\sqrt{2}}\cos\phi\sin\theta,\frac{1}{\sqrt{2}}
\sin\phi\sin\theta,\frac12(s_{\theta}+\cos\theta)\right]
 \label{N104}\end{equation}
for the optimal PCC, where $s_{\theta}={\rm sgn}(\pi-2\theta)$.
Actually, $s_{\theta}$ for $\theta=\pi/2$ can take any value in
$[-1,1]$. This is because for $\theta=\pi/2$ any linear
combination of the optimal PCC transformations for south
hemisphere ($|00\>\rightarrow|00\>$, $|10\>\rightarrow |\psi_+\>$)
and north hemisphere ($|00\>\rightarrow|\psi_+\>$,
$|10\>\rightarrow |11\>$) is also optimal \cite{F1}.

\begin{figure}[t!]
\hspace*{-3mm} \epsfxsize=7cm \epsfbox{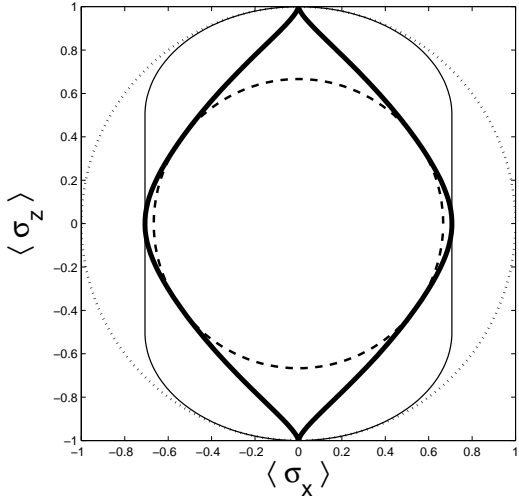}
\caption[]{Cross sections of Bloch representation for the optimal
cloners: the MPCC (thick solid), the PCC (thin solid), the UC
(dashed), and unphysical perfect cloner (dotted
curve).}\label{fig3}
\end{figure}

\section{Optimality proof for the MPCC}

Let $\lambda=\tr_{\rm out}(R\chi)$ be the matrix of Lagrange
multipliers, then
\begin{equation} \label{N20}
\lambda=\frac{1}{4}[(1+c_1^2)A+2B+2s_1^2C]\,\openone_{\rm in}.
\end{equation}
To prove that the derived $\chi$ is the optimal TPCP map one
should show that (i) $\Delta=\lambda\otimes\openone_{\rm out}-R$
is a positive semidefinite matrix, and (ii) $\tr \lambda$
saturates the fidelity bound, i.e., $F\le\tr\lambda$ \cite{F2}.
From a technical point of view, it is more convenient to prove
condition (ii) first.

Using Eqs. (\ref{N14}), (\ref{N17}), and (\ref{N20}) we derive
\begin{eqnarray}\nonumber
\tr\lambda-F&=&\frac{1}{2}(1+c_1^2)A+B+s_1^2C\\\nonumber
&&-\frac{1+\Lambda^2}{2}+\frac{1}{2}s_1^2\Lambda(\Lambda-\bar{\Lambda}\sqrt{2})\\\nonumber
&=&\frac{\Lambda^2}{4}\left[\left(1+c_1\right)^2+\left(1-c_1\right)^2-2c_1^2-2
\right]\\ \label{N21} &=&0.
\end{eqnarray}
This implies that the Eq. (\ref{N19}) can be rewritten as
\begin{equation}\label{N22}
\lambda=\frac{F}{2}\,\openone_{\rm in}.
\end{equation}
The eigenvalues of $\Delta$ can be written in terms of fidelity
and $R$ matrix elements as follows:
\begin{eqnarray}
\nonumber
\delta_1&=&\frac{1}{2}\left(F-\frac{1}{2}\right),\\\nonumber
\delta_2&=&\frac{1}{2}\left(F-\frac{s_1^2}{2}\right),\\
\delta_{3,4}&=&\frac{1}{2}\left(F-R_{11}-R_{22}\pm\bar
R\right),\label{N22a}
\end{eqnarray}
where $\bar R^2=(R_{11}-R_{22})^2+8R_{16}^2$. All the eigenvalues
are double degenerate. Moreover, the following equation is
satisfied:
\begin{equation}\label{N23}
F=R_{11}+R_{22}+\bar R.
\end{equation}
Therefore, $\delta_3=F-(3-s_1^2)/4$ and $\delta_4=0$. Since
$F>3/4$ we see that $\forall_i \delta_i\ge0$, and hence $\Delta$
is positive semidefinite. This statement completes the proof.

\section{Implementations of the optimal MPCC}

We propose two quantum circuits, shown in Figs. 4 and 5, which
implement the optimal MPCC by transforming the input state
$|\psi_{\rm in}\rangle=a|000\rangle+b|100\rangle$ into
\begin{equation}
|\psi_{\rm out}\rangle=a(\Lambda|000\rangle
+\bar{\Lambda}|\psi_{+}\rangle|1\rangle)
+b(\Lambda|111\rangle+\bar{\Lambda}|\psi_{+}\rangle|0\rangle).
\label{N101}
\end{equation}
The quantum circuit depicted in Fig. 4 performs the following
transformation:
\begin{equation}
|\psi_{\rm out}\rangle =
 U_{\rm CNOT}^{(32)}U_{\rm CNOT}^{(21)}U_{\rm CNOT}^{(13)}
 U^{(32)}_{\rm CH} R^{(3)}_y(\gamma_{\theta}) |\psi_{\rm in}\rangle,
 \label{N101b}
\end{equation}
where the superscripts indicate qubits for which the corresponding
gate is applied. The basic element of the circuit is the rotation
\begin{equation}
R_y(\gamma_{\theta})=\left[\begin{array}{cc}
\cos(\gamma_{\theta}/2) & -\sin(\gamma_{\theta}/2)\\
\sin(\gamma_{\theta}/2) & \cos(\gamma_{\theta}/2)\\
\end{array}\right],
\label{N101c}
\end{equation}
about $y$ axis for angle $\gamma_{\theta}=2\arccos
\Lambda({\theta})$. In addition, this circuit is composed of controlled NOT (CNOT) gates, $U_{\rm CNOT}$, and controlled Hadamard gate, $U_{\rm CH}$,
which can be decomposed as $U^{(32)}_{\rm CH}=A^{(2)}U^{(32)}_{\rm
CNOT}A^{(2)}$, where
\begin{equation}
A=\frac{1}{\sqrt{4+2\sqrt{2}}}\left[\begin{array}{cc}
1 & 1+\sqrt{2}\\
1+\sqrt{2} & -1\\
\end{array}\right].
\label{N101d}
\end{equation}
\begin{figure}[t!]
\hspace*{-3mm} \epsfxsize=9.5cm \epsfbox{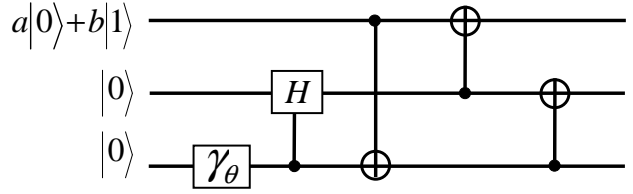} \caption[]{A
quantum circuit which implements the optimal MPCC. From left to
right: rotation $R_y(\gamma)$ about $y$-axis, controlled Hadamard
gate $U_{\rm CH}$, and three CNOT gates.}\label{fig4}
\end{figure}

The basic idea behind the second quantum circuit, shown in Fig. 5,
is different. The circuit acts as follows. The two CNOT gates in
Fig. 5 implement the standard repetition code, which, together with
the NOT gate applied to third qubit, lead to
$|\psi_{1}\rangle=\sigma_{x}^{(3)}U_{\rm CNOT}^{(13)}U_{\rm
CNOT}^{(12)} |\psi_{\rm in}\rangle=a|001\rangle+b|110\rangle$. In
our implementation, the basic gate $U(t)$ gate corresponds to the
evolution operator $\exp(-iHt)$ of a system described by the
interaction Hamiltonian
\begin{equation}
H=\frac{\hbar\kappa}{2}\sum_{n\neq
m=1}^{3}\sigma_{+}^{(n)}\sigma_{-}^{(m)}+\sigma_{-}^{(n)}\sigma_{+}^{(m)},
\label{N102}
\end{equation}
where $\kappa$ is an effective coupling constant and
$\sigma_{\pm}^{(n)}=\sigma_{x}^{(n)}\pm i\sigma_{y}^{(n)}$ are,
respectively, the spin raising and lowering spin operators acting
on the $n$th qubit. We refer to the system, described by Eq.
(\ref{N102}), as the equivalent-neighbor model \cite{Miran}.
General solution of the Schr\"odinger equation for Hamiltonian
(\ref{N102}) is known assuming any number of qubits and arbitrary
initial conditions (see, e.g., \cite{Miran}). In our special case
of three qubits, one has
\begin{eqnarray}
  U(t)|001\> = C^{31}_0(t)|001\>+C^{31}_1(t)(|010\>+|100\>), \nonumber \\
  U(t)|110\> = C^{32}_0(t)|110\>+C^{32}_1(t)(|011\>+|101\>),
\label{N109}
\end{eqnarray}
where
\begin{eqnarray}
  C^{3M}_0(t) &=& \frac13 (e^{-2i\kappa t}+2e^{i\kappa t}),
\nonumber \\
  C^{3M}_1(t) &=& \frac23 \sin\left(\frac32 \kappa t\right)e^{-\left(\frac{i}2\right)(\pi+\kappa t)}
\label{N114}
\end{eqnarray}
for both "excitation" numbers $M=1,2$. Note that
$|C^{3M}_0(t)|^2+2|C^{3M}_1(t)|^2=1$.  Let us choose such that
 evolution
time $t=t_\theta$ that $\bar\Lambda=\sqrt{2}|C^{31}_2(t)|$ for a
given $\theta$. Thus, after interaction time
\begin{eqnarray}
  t_\theta &=& \frac{2}{3\kappa} {\arcsin} \left(\frac3{2\sqrt{2}}\,
  \bar\Lambda\right),
\label{N105}
\end{eqnarray}
state $|\psi_{1}\rangle$ is transformed into
\begin{eqnarray}
|\psi_{2}\rangle=U(t_\theta)|\psi_{1}\rangle= e^{i\varphi_1}\big[
a(\Lambda e^{i\varphi/2}|001\rangle
+\bar{\Lambda}|\psi_{+}\rangle|0\rangle)
\nonumber \\
 +b(\Lambda e^{i\varphi/2} |110\rangle+\bar{\Lambda}
 |\psi_{+}\rangle|1\rangle)\big], \label{N106}
\end{eqnarray}
where $\varphi_k={\rm arg}[C^{31}_k(t_\theta)]$ ($k=0,1$) and
$\varphi=2(\varphi_0-\varphi_1)$. The global phase factor
$\exp(i\varphi_1)$ is physically irrelevant and can be dropped.
While the relative phase factor $\exp(i\varphi/2)$ can be
corrected by applying the double-controlled rotation gates (see
Fig. 5):
\begin{eqnarray}
  U_{\rm CCR}(\varphi) = I_8 + f(-\varphi)|110\>\<110| +
  f(\varphi)|111\>\<111|,
\nonumber \\
  U_{\rm ccR}(-\varphi)  = I_8 + f(\varphi)|000\>\<000|
  +f(-\varphi)|001\>\<001|, \nonumber
\label{N107}
\end{eqnarray}
which generalize the standard $Z$-rotation gate $R(\varphi)={\rm
diag}\{[\exp(-i\varphi/2),\exp(i\varphi/2)]\}$. In the above equations,
$f(\varphi)=\allowbreak \exp(i\varphi/2)-1$ and $I_N$ is the
$N\times N$ identity matrix. So finally, it is seen that
$|\psi_{\rm out}\rangle\sim \sigma_{x}^{(3)} U_{\rm
ccR}(-\varphi)U_{\rm CCR}(\varphi)|\psi_2\>$ is the desired cloned
state, given by Eq. (\ref{N101}).

\begin{figure}[t!]
\hspace*{-3mm} \epsfxsize=9.5cm \epsfbox{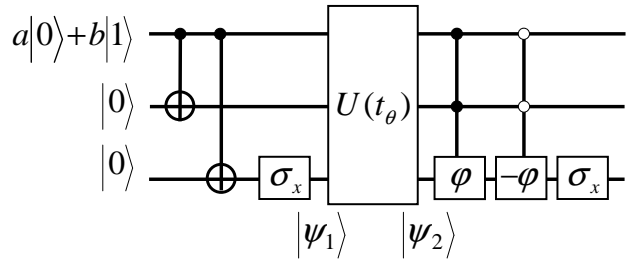}
\caption[]{Another quantum circuit implementing the optimal MPCC.
The double-controlled gate with $\varphi$ ($-\varphi$), in the
text denoted by $U_{\rm CCR}$ ($U_{\rm ccR}$), corresponds to
conditional rotation by a phase $\pm \varphi$, where the
conditions, marked by the filled (empty) circles, are satisfied if
both qubits 1 and 2 have the bits in the 1 (0) states. The first
two gates are the CNOT gates, $\sigma_x$ is the NOT gate, and the
action of gate $U(t)$ corresponds to the evolution within time
$t=t_\theta$ of the equivalent-neighbor system described by
Hamiltonian (\ref{N102}). }\label{fig5}
\end{figure}

The equivalent-neighbor model of three qubits can be implemented in
various ways. Here, we shortly discuss a quantum-dot model
proposed by Imamo$\rm\bar{g}$lu {\em et al.} \cite{Imamoglu}. The model
describes interaction of $N$ quantum-dot spins mediated by a
single-mode cavity and laser fields. In our case, three
semiconductor quantum dots, each with a localized single
conduction-band electron, embedded inside a microdisk structure
and addressed selectively by laser beams of frequency $\omega
_{n}^{(L)}$ ($n=1,2,3$) and intensity $|E _{n}^{(L)}|^2$ to induce
strong coupling of the electron spins to a single cavity mode of
frequency $\omega _{{\rm cav}}$. The energy spectrum of each dot
is represented by three states: $|0\rangle_n$ and $|1\rangle_n$
representing the conduction-band electron spin states, while the
effective state $|v\rangle_n$ describes all valence-band levels of
the $n$th dot. It can be shown \cite{Imamoglu,Miran}, by applying
the adiabatic elimination method of the valence-band levels and
the cavity field, that the system can be described by the
following effective interaction Hamiltonian:
\begin{eqnarray}
\hat{H}_{{\rm eff}}=\!\frac{\hbar}{2}\sum_{n\neq m}\kappa
_{nm}(t)\left[{\hat{\sigma}}
_{n}^{+}{\hat{\sigma}}_{m}^{-}e^{i(\Delta _{n}-\Delta _{m})t} +
{\rm h.c.}\right],\label{N202}
\end{eqnarray}
where $\kappa _{nm}(t)$ is the effective two-dot coupling strength
between the electron spins in the $n$th and $m$th dots being a
function of frequencies, detunings (see Fig. 1 in \cite{Miran}),
intensities $|E _{n}^{(L)}|^2$, and the dipole coupling strengths
between the levels $|0\rangle_n$ and $|1\rangle_n$, and $|v\rangle_n$. The effective Hamiltonian
apparently describes direct coupling between the spins of the
$n$th and $m$th dots, but it should be stressed that in the real
microscopic picture the coupling between the spins is only
indirect via the cavity and laser fields. To realize the
equivalent-neighbor model it is enough to ensure (i) the same
detunings $\Delta _{n}={\rm const}$ for all dots by adjusting the
laser-field frequencies $\omega _{n}^{(L)}$ and (ii) the same
effective coupling constants $\kappa _{nm}(t)={\rm const}$ by choosing
the proper laser intensities, $|E^{(L)}_n|^2$ and $|E^{(L)}_m|^2$
for the adequate frequencies $\omega _{n}^{(L)}$ and $\omega
_{m}^{(L)}$. Thus, Hamiltonian (\ref{N202}) reduces to Eq.
(\ref{N102}). We note that other implementations (see, e.g.,
\cite{Reina}) of the equivalent-neighbor model of interacting
quantum dots can be applied here.

Imamo$\rm\bar{g}$lu {\em et al.} \cite{Imamoglu} described how to perform
the conditional phase flip between $m$th and $n$th dots
gates, which combined with single-qubit rotations can be used to
implement the CNOT gates. Analogously, the CCR gates can be
realized in their model.

To implement of the $U_{\rm CCR}$ and $U_{\rm ccR}$ gates, we
recall the well-known theorem in quantum information that any
three-qubit controlled gates can be replaced by two-qubit
controlled gates. Specifically, by applying lemma 6.1 of Barenco
{\em et al.} \cite{Barenco}, one gets $ U_{\rm CCR}(\varphi) =
U_{\rm CR}^{(13)}(\varphi/2) U_{\rm CNOT}^{(12)} U_{\rm
CR}^{(23)}(-\varphi/2) U_{\rm CNOT}^{(12)} U_{\rm
CR}^{(23)}(\varphi/2)$, which is given in terms of the CNOT gates
and two-qubit controlled $Z$-rotation gates $U_{\rm CR}(\varphi) =
I_4 + f(-\varphi)|10\>\<10| + f(\varphi) |11\>\<11|$. Moreover,
$U_{\rm ccR}(-\varphi)$ can be replaced by
$\sigma_{x}^{(1)}\sigma_{x}^{(2)}U_{\rm
CCR}(-\varphi)\sigma_{x}^{(1)}\sigma_{x}^{(2)}$.

\section{Conclusion}

We proposed a  $1\rightarrow 2$ quantum cloning machine of
an input qubit state assuming {\em a priori} information of
$|\langle \sigma_z\rangle|$ (or, equivalently, $\sin\theta$). We
refer to this machine as the mirror phase-covariant cloner by
contrast to the standard phase-covariant cloner of an input state
with {\em a priori} information of $\langle \sigma_z\rangle$ (or
$\theta$). We found analytical expressions describing the cloning
transformation and fidelity of the clones. Applying a generalized
method of Fiur\'a\v{s}ek \cite{F1}, we derived and solved extremal
equations for the optimal cloning transformation, which provides
lower fidelity than that for the optimal phase-covariant cloner
\cite{F2}. This is because our transformation was derived assuming
less knowledge about the state to be cloned. Nevertheless, our
cloning machine for the whole range of $\theta$ provides fidelity
higher $\left( { \rm except} \theta=\pi/2\pm[\pi/2-{\rm acos}(\sqrt3/3)] \right)$ than
the fidelity $F=5/6$ of the universal cloner
\cite{Buzek98,Bruss98}. The fidelity of those optimal cloning
machines (see Figs.~2 and 3) can be used as thresholds for secure
quantum communication and quantum teleportation \cite{Sahin}.
Finally, we proposed quantum circuits as an implementation of the
optimal cloning transformation and suggested a physical
realization in a quantum-dot model Imamo$\rm\bar{g}$lu {\em et al.}
\cite{Imamoglu}.

\section*{ACKNOWLEDGMENTS} 
We thank Jaromir Fiur\'a\v{s}ek and Yu-xi
Liu for discussions. \c{S}.K.\"O. was partly supported by MEXT
Grant-in-Aid for Scientific Research on Innovative Areas Grant No. 20104003.
The research was conducted within the LFPPI network.

\end{document}